\documentclass[preprint,5p]{elsarticle}

\usepackage{xpatch}
\usepackage{xpatch}
\makeatletter
\xpatchcmd{\pprintMaketitle}{\hrule}{}{}{}
\xpatchcmd{\pprintMaketitle}{\hrule}{}{}{}

\xpatchcmd{\MaketitleBox}{\hrule}{}{}{}
\xpatchcmd{\MaketitleBox}{\hrule}{}{}{}

\xpretocmd{\pprintMaketitle}{\vspace*{-1cm}}{}{}
\xpretocmd{\MaketitleBox}{\vspace*{-1cm}}{}{}
\makeatother

\makeatletter
\def\ps@pprintTitle{%
  \let\@oddhead\@empty
  \let\@evenhead\@empty
  \let\@oddfoot\@empty
  \let\@evenfoot\@empty}
\makeatother

\makeatletter
\xpatchcmd{\@address}{\addsep\footnotesize\itshape}{\addsep\normalsize\itshape}{}{}
\makeatother

\makeatletter
\xpatchcmd{\@address}{\par\vskip6pt}{\par\vskip1pt}{}{}
\xpatchcmd{\@address}{\par\vskip8pt}{\par\vskip2pt}{}{}
\makeatother

\makeatletter
\renewenvironment{abstract}{%
  \onecolumn
  \global\setbox\absbox=\vbox\bgroup
  \vspace*{-4.5em}
  \begin{center}\bfseries Abstract\end{center}%
  \medskip
  \leftskip=0.12\textwidth
  \rightskip=\leftskip
  \noindent\ignorespaces
}{%
  \par\egroup
  \twocolumn
}
\makeatother

\usepackage{tikz}
\usepackage{xcolor}
\usepackage{graphicx}
\usepackage{lineno,hyperref}
\usepackage{amsmath}
\usepackage{amssymb}
\usepackage{subcaption}
\usepackage{algpseudocode}
\usepackage{algorithm}
\modulolinenumbers[10]

\usepackage{lineno}

\newcommand{\R}{\mathbb{R}}

\newcommand{\N}{\mathbb{N}}

\usepackage{graphicx}

\newcommand{\xlim}[2]{\underset{#1\to#2}\lim}

\renewcommand{\d}[2]
{\frac{d#1}{d#2}}

\newcommand{\nc}{\newcommand}
\nc {\Prod}{(\underset{I}\phi\, r_I^{n_I})}
\nc {\wlw}{\wedge\ldots\wedge}
\nc {\olo}{\otimes\ldots\otimes}
\nc{\bea}{\begin{eqnarray*}}
\nc{\eea}{\end{eqnarray*}}
\nc {\e}{\varepsilon}
\nc{\de}{\delta}
\nc{\A}{\hat a}
\nc{\Ad}{\hat a^\dag}
\nc{\grad}{\nabla}
\nc{\nlim}{\underset{n\to\infty}\lim}
\nc{\ilim}{\underset{i\to\infty}\lim}
\nc{\isum}{\sum\limits_{i=1}^N}
\nc{\xx}{\underset{x\to x_0}\lim}
\nc{\Bnrm}{\Big |\Big|}
\nc{\sym}{\text{Sym}}
\nc{\inte}{\overset{\circ}}
\nc{\del}{\partial}
\nc{\plp}{+\ldots+}
\nc{\lre}{\longrightarrow}
\nc{\be}{\begin{equation}}
\nc{\ee}{\end{equation}}
\nc{\CP}{\mathbb{CP}}
\nc{\End}{\text{End}}
\nc{\mf}{\mathfrak}
\nc{\Hom}{\text{Hom}}
\nc{\spec}{\text{Spec}}
\nc{\sub}{\subseteq}
\nc{\weakto}{\rightharpoonup}
\nc{\ph}{\varphi}
\nc{\leqc}{\lesssim}
\nc{\delbar}{\overline \del}
\nc{\Tr}{\text{Tr}}
\nc{\Lhe}{\mathcal L_{(\phii^{h_\e}, A^{h_\e})}}

\begin{document}
\vspace{-1cm}
\begin{frontmatter}

\title{A Propagator-based Multi-level Monte Carlo Method for Kinetic Neutral Species in Edge Plasmas}
\author[ucb]{Gregory J. Parker\corref{cor1}}
\cortext[cor1]{\noindent Corresponding author. Email address: \texttt{gjparker@berkeley.edu}}

\author[llnl]{Maxim V. Umansky}
\author[llnl]{Benjamin D. Dudson}

\address[ucb]{University of California, Berkeley, Department of Mathematics, Berkeley CA 94720, USA}
\address[llnl]{Lawrence Livermore National Lab, Livermore, CA 94550, USA}

\begin{abstract}
We propose and investigate a new multi-level Monte Carlo scheme for numerical solutions of the kinetic Boltzmann equation for neutral species in edge plasmas. In particular, this method explicitly exploits a key structural property of neutral particle dynamics: the prevalence of frequent collisions for which the outgoing velocity is determined by local plasma parameters. Using this property, we derive a multi-level algorithm based on collision event propagator and show, both analytically and through numerical experiments, that it reproduces the results of standard Monte Carlo methods. We further demonstrate that, in the context of coupled plasma-neutral edge simulations employing correlated Monte Carlo, the proposed scheme retains trajectory correlation to machine precision as the system evolves, whereas conventional methods exhibit rapid decorrelation. These results indicate that the propagator-based multi-level Monte Carlo scheme is a promising candidate for use in fully implicit Jacobian-free Newton-Krylov (JFNK) solvers for coupled plasma-neutral systems. 
\vspace{.5cm}
\end{abstract}

\end{frontmatter}

\section{Introduction}

\vspace{-.2cm}

The edge of magnetic fusion devices hosts a quagmire of complex physical effects, and the need for accurate modeling of these effects constitutes a key barrier in the quest for usable fusion energy \cite{stangeby2000plasma,militello2022boundary,krasheninnikov2020edge}. In the edge region, the fusion plasma interacts strongly with neutral atomic species, and runaway non-linear cascades arising from these interactions can cause chemical sputtering, undermine divertor detatchment, and generate excessive heat loads to the device walls \cite{wiesen2017plasma,reiter1992progress}.  As a result, predictive edge modeling demands a quantitatively faithful description of the dynamics of the coupled plasma-neutral system.

The modeling of neutral species in particular remains a challenging computational task. Accurate prediction of these species generally requires numerically solving the full kinetic Boltzmann equation. Indeed, neutral species can occupy vast regions of the effective parameter space and will not, in general, remain within the domain of validity of any reduced model across different regions of the device, different times in a discharge, and different plasma scenarios. The high-dimensional and non-local nature of the full Boltzmann equation, the complex geometries involved, and immense numbers of atomic species present conspire to make solving the full Boltzmann equation a daunting task, even on world-class high-performance computing clusters.

At present, Monte Carlo (MC) is, in many cases, the most robust method for neutral modeling due to its advantageous scaling with the dimension of the phase space, and because it makes no {\it a priori} assumptions about the parameter regime that neutral species occupy \cite{heifetz1986neutral,wiesen2015new,stotler1994neutral}. Additionally, Monte Carlo enjoys the advantages of easy implementation of complex atomic physics, and linear scaling with the number of neutral species, whereas continuum models are generally quadratic. 

Despite these advantages, Monte Carlo remains a computationally costly method, and its stochastic nature introduces additional algorithmic challenges when coupled to plasma models \cite{dimits2017efficient,joseph2017coupling}. Standard Monte Carlo methods average large ensembles of independently sampled particle trajectories, resulting in poor compatibility and slow convergence when coupled to modern fully implicit plasma solvers. Indeed, many large-scale edge plasma simulations require weeks to months of wall computation time to reach a steady-state of the coupled system \cite{kotov2014b2,hatzky2012hlst, boeyaert2024numerical,lore2023time}. Jacobian-free Newton-Krylov (JFNK) solvers, which are now common in large-scale continuum plasma models \cite{dudson2009bout++,rognlien2002application}, have the potential to greatly reduce these immense computational costs for coupled systems. Even for correlated Monte Carlo, however, rapid decorrelation of particle trajectories precludes the formation of stable Jacobian-vector products, and thus the use of Jacobian-free Newton-Krylov (JFNK) \cite{dimits2017efficient,joseph2017coupling,parker2025coupling,willert2015newton}. These challenges motivate the development of new kinetic algorithms that retain the fidelity of the full Boltzmann description while opening the possibility of Jacobian-based methods. 

The purpose of this article is to propose such an algorithm and to give preliminary demonstrations of the resulting multi-level Monte Carlo scheme in coupled-plasma neutral simulations. The basis of the new method is the following property of neutral particle dynamics: 

\newpage 
\begin{itemize}
\item[(VLP)] Particles undergo frequent collisions for which the outgoing velocity is determined by local plasma parameters. 
\end{itemize}

\noindent We refer to (VLP) as the {\it velocity-localizing property}. For neutral particles, resonant charge-exchange collisions respect this property to a high-degree of accuracy in most regimes \cite{DEGAS2manual}. In many standard Monte Carlo methods, this property manifests as the ``velocity-swap'' procedure, by which the outgoing neutral particle takes on the velocity of a thermal ion of the same species at the collision \cite{heifetz1986neutral,wiesen2015new,DEGAS2manual}. 

Mathematically, (VLP) implies that in a snapshot of particle trajectories infinitesimally after a fixed number of resonant charge-exchanges, the neutral particle distribution $n(x,v)$ on phase space collapses to a product distribution:  

\be n(x,v) \lre f(x)\mathcal N_{\mathfrak p(x)}(v)\ee

\noindent wherein $\mathcal N_{\frak p(x)}(v)$ is fixed velocity distribution depending on the local plasma state $\frak p(x)$ at $x$. In the ``velocity-swap'' setting, $\mathcal N(v)$ is a Maxwellian determined by the local plasma temperature and velocity. The algorithm introduced in the present work uses this collapse to compress particle evolution operators into propagator matrices of the {\it real-space} grid dimension, from which the equilibrium distribution can be recovered via linear algebra. This method has some precedent in the propagator-based method of Valanju and LaBombard in the continuum setting \cite{valanju1990nut,bombard2001kn1d}, and some relation to fission-matrix Monte-Carlo methods in neutron transport \cite{carney2014theory,walters2018rapid}, although (VLP) does not hold in that setting.  

The remainder of the article is organized as follows. The mathematical foundations of the new approach are presented in Sections \ref{kernelssection}--\ref{MLMCSection}. Section \ref{PropagatorMLMC} describes the numerical implementation of the new method in a two-dimensional setting with simplified atomic physics on rectilinear tokamak geometries. Numerical results are presented in Sections \ref{reproducessection}--\ref{correlatedcoupled}. Section \ref{reproducessection} confirms that the new approach reproduces the same equilibrium distribution as standard methods, and Section  \ref{correlatedcoupled} displays the results of coupled plasma-neutral simulations. The discussion and conclusion occupy the final two sections.

\section{Kernel Methods for the Boltzmann Equation}
\label{kernelssection}
For a region $\Omega\subseteq \R^d$ for $d=1,2,3$ with boundary $\del\Omega$, our focus is on numerical solutions of the kinetic Boltzmann equation 
\be \d{n(x,v)}{t} + v\cdot \grad n(x,v)  + C_{\frak p}(n(x,v)) = s_0(x,v) \label{Boltz}\ee
\noindent for the density of a neutral species $n(x,v)$ on the phase space $T^*\Omega=\Omega\times \R^d$, subject to boundary conditions along the boundary $\del\Omega \times \R^d$. Here, $C_{\frak p}$ is a linear collisional integral operator incorporating various interactions, which depends non-linearly on a vector-valued background plasma parameter field $\frak p(x)$, and $s_0(x,v)$ is an external source term. 

In practice, $\Omega$ is typically a two-dimensional cross-section of a tokamak or stellarator, $s_0(x,v)$ is the neutral source from the device walls, and $C_\frak p$ typically includes the following interactions: (i) charge-exchange, (ii) ionization, and (iii) recombination of neutral atoms, and sometimes (iv) collisions with a background neutral mean-field. The plasma parameter $\frak p$ can include e.g. ion/electron density, temperatures, and velocity. Although the method described below can accommodate all the interactions (1)--(4), all numerical experiments in the presented in this study include only (1)--(2) for simplicity. 

Because Eq. (\ref{Boltz}) is linear in $n(x,v)$, there is a steady-state Green's function $K(x,v,x',v'; \frak p)$ which solves the time-independent version of (\ref{Boltz}) with a point source $s_0(x,v)=\delta(x-x,v-v')$. For a general source, the solution to (1) is given by the convolution \be n_\star(x,v)=\int_{T^*\Omega}  K(x,v,x',v'; \frak p)s_0(x',v')  \ dx' dv' \label{convolution}  \ee

\noindent in the standard way. In particular, knowledge of the (family of) Green's functions $K(x,v,y,v';\frak p)$ is sufficient to construct the steady-state distribution, by linearity. The interpretation of $K(x,v,x',v';\frak p)$ is that it is the probability density that an infinitesimal volume element at $(x,v)$ receives a particle from a point-source at $(x',v')$ after the system has relaxed to equilibrium. Henceforth we will use $z=(x,v)$ to denote phase space variables, thus we write $K(z,z';\frak p)$, and use $K_1*K_2=\int  K_1(z,z')K_2(z',z'') \ dz' $ to denote the convolution.

\subsection{Kernels with the Semi-Group Property}
\label{semigroupprop}
A family of kernels \footnote{By ``kernel'' we mean a function on $T^*\Omega\times T^*\Omega$, which induces an operator by convolution. A Green's function is the specific kernel for which the induced operator is the solution operator of Eq. \ref{Boltz}, and a Propagator is one such that the induced operator advances the state by a particular parameter. } $K_\sigma(z,z';\frak p)$    parameterized by a real number $\sigma>0$ is said to have the {\bf semi-group property} if they satisfy the following criteria: 

\begin{enumerate}
\item[(A1)] ${\displaystyle   K_\sigma(z,z';\frak p)\  * \ K_{\tau}(z',z'';\frak p)=K_{\sigma+\tau}(z,z'';\frak p),}$
\medskip 
\item[(B1)] $\xlim{\sigma}{\infty} K_{\sigma}(z,z';\frak p)=K(z,z';\frak p),$
\end{enumerate}
\noindent where the unsubscripted $K(z,z';\frak p)$ is the steady-state Green's function as in Eq. (\ref{convolution}).  The canonical example of a family kernels with the semi-group property are the time-evolution propagators $K_t(z,z';\frak p)$ for $t\geq 0$ which are the Green's function for evolution by time $t$, though as we will see, this is not the only useful family.

A standard mathematical result is the following: {\it If $K_\sigma$ is a family of kernels obeying the semi-group property (A1)--(B1), then the equilibrium distribution can be obtained by repeated convolution.} Indeed, given any fixed $\sigma>0$, the equilibrium distribution of Eq. \ref{Boltz} is obtained as the limit 
\begin{eqnarray}
n_\star(z)&=& K(z,z';\frak p) \ * \ s_0(z') \nonumber \\
&=& \xlim{M}{\infty} K_{M\sigma}(z,z';\frak p) \ * \ s_0(z') \label{repconv}\\
&=&\xlim{M}{\infty} \underbrace{K_\sigma * K_\sigma *\ldots K_\sigma * s_0}_{\text{$M$ times}} \nonumber
\end{eqnarray} 
\noindent where the two lines follow from invoking properties (B1) and (A1) respectively. More generally, one does not need $\sigma$ to be parameterized continuously, only for it to be an additive semi-subgroup of the real numbers, hence the name. 

Our setting will require a mild generalization of the standard semi-group property (A1)--(B1), which appears as (A2)--(C2) below. This generalization covers the case that the semi-group property does not hold directly for a family of kernels $K_\sigma$ that yield the equilibrium distribution, but instead holds for an auxiliary family of kernels $J_\sigma$ that yield a related quantity of interest from which the equilibrium can be reconstructed. Thus (A2)--(C2) make reference to two families of kernels $J_\sigma$ and $K_\sigma$ for $\sigma>0$. In our particular setting, $J_\sigma$ will be the {\it source-correction operator}, which effectively advances the particles' current location towards equilibrium position (where fluxes are balanced), and $K_\sigma$ will be any standard Monte Carlo estimator (e.g. a track-length estimator). The former tallies particles' current locations, whereas the latter tallies where they have spent time. These quantities are, in general, completely independent\footnote{Consider the example of a thermal bath of particles in a box with a very small puncture and a source replenishing them at exactly the rate they escape. Here, the source correction operator $J_\sigma$ advances the current location of the particles, and as $\sigma\to \infty$ each particle will exit the box with probability $1$. Thus in this case the current particle location distribution function approaches zero in the interior of the box, and the additional quantity of where particles have spent time along the way is needed to reconstruct the equilibrium distribution, this being precisely what $K_\sigma$ tallies. }.

Two families of kernels $J_\sigma(z,z';\frak p)$ and $K_\sigma(z,z';\frak p)$ parameterized by a real number $\sigma$ in an additive semi-group of $\R^{\geq 0}$ are said to obey the {\bf generalized semi-group property} if they satisfy
\medskip 

   \begin{enumerate}
\item[(A2)] ${\displaystyle   J_\sigma(z,z';\frak p)\  * \ J_{\tau}(z',z'';\frak p)=J_{\sigma+\tau}(z,z'';\frak p),}$
\medskip 
\item[(B2)] $ \xlim{\tau}{\infty} K_{\tau}(z,z';\frak p)=K(z,z';\frak p)$
\item[(C2)] $ K_{\sigma+\tau}(z,z';\frak p)*s_0 = K_\sigma(z,z';\frak p) * (J_\tau(z,z';\frak p)*s_0)$.
\end{enumerate}

\medskip

\noindent when $\tau=M\sigma$ is now constrained to be a positive integer multiple of $\sigma$. Note that we no longer assume (A1). In special settings where $K_\sigma=J_\sigma$ coincide, then (A2) and (C2) collapse to become (A1), and the generalized semi-group property reproduces the previous one.

In this more general setting, the correct adaptation of the repeated convolution process Eq. (\ref{repconv}) is as follows. For any chosen $\tau>0$, the equilibrium may now be obtained as the limit 
\begin{eqnarray}
n_\star(z)&=&\xlim{M}{\infty} K_\tau(z,z';p) (J_{M\sigma}*s_0)\label{internalkernel}\\
&=& \xlim{M}{\infty} K_\tau(z,z';p) (\underbrace{J_{\sigma}*J_{\sigma}*\ldots  *J_{\sigma}}_{M \text{ times}}*s_0).\nonumber
\end{eqnarray}

An equivalent, perhaps more intuitive way to describe Eq. (\ref{internalkernel}) is to introduce the {\bf virtual source term}

 \be \label{sstardef}s_\star =\xlim{M}{\infty}(\underbrace{J_{\sigma}*J_{\sigma}*\ldots  *J_{\sigma}}_{M \text{ times}}*s_0).\ee

 \noindent Substituting into Eq. (\ref{internalkernel}) with $\tau=\sigma$, this definition shows that $s_\star$ satisfies \be K_\sigma(z,z';\frak p)*s_\star(z)  \ =  \ n_\star(z) \ = \  K(z,z';\frak p)* s_0(z), \label{finalapplication}\ee

 \noindent where the second equality is Eq. (\ref{convolution}). Therefore, in the case that convolution with $K_\sigma$ is invertible (which is generally the case in practice) $s_\star$ satisfies  

 \be s_\star(z)=(K_\sigma(z,z';\frak p)* )^{-1}K(z,z';\frak p)*s_0(z).\label{sstarsolves}\ee
 
 \noindent Thus the virtual source term may be interpreted as precisely giving the distribution the particles {\it would have to have} in order for a single application of $K_\sigma$ to produce the equilibrium. The intermediate iterates $J_{M\sigma}*s_0$ may be interpreted as source terms only partially corrected toward the virtual source, which explains the name of $J_\sigma$ as the source-correction operator. 
 
With this interpretation, (A2)--(C2) may be viewed as a two-step process by which one first produces the virtual source $s_\star$ by repeated convolution with $J_\sigma$, and then reconstructs the equilibrium $n_\star$ from $s_\star$ by a single, final application of $K_\sigma$. 

\subsection{Linear Algebra for the Virtual Source}
\label{multilevel}
Given $J_\sigma$ and $K_\sigma$, the repeated convolution scheme Eq. (\ref{internalkernel}) allows the equilibrium distribution to be obtained by linear algebra in two different, albeit related, ways.  Write the convolution operator as \be J_\sigma(z,z';\frak p) * s  =(\text{Id} + P_\sigma)s.\label{sourcecorrection}\ee

\noindent We refer to the operator $P_\sigma$ as the propagation operator or simply the {\bf propagator} and omit the dependence on $\frak p$ from the notation. In practice, both for time-stepping and the family of event-based kernels, any non-trivial amount of ionization implies that $\|P_\sigma\|<1$ in operator norm, hence $(\text{Id}+P_\sigma)$ is invertible with convergent Neumann series. 

\subsubsection{Linear System} \label{linsystem} The first method to obtain $s_\star$, and thus the equilibrium distribution, is by directly solving the linear system defining it. By Eq. (\ref{sstardef}) and Neumann series,  \begin{eqnarray} s_\star=\xlim{M}{\infty} (\text{Id}+P_\sigma)^M s_0 = (\text{Id}-P_\sigma)^{-1}s_0  \label{sstardef}\end{eqnarray}

\noindent so $s_\star$ solves the linear system \be (\text{Id}-P_\sigma)s_\star=s_0. \label{linsstar}\ee

\noindent In a chosen discretization of Eq. (\ref{Boltz}), this is a finite-dimensional invertible linear system. 

In the case where (A1)--(B1) hold directly (which is the case when using a collisional estimator), then the situation is simpler since $K_\sigma=J_\sigma$. In this case, the equilibrium distribution can be obtained as $n_\star(x,v)= s_\text{cx} \cdot \frac{1}{n_i K_{\text{cx}}}$ where $n_i$ is the ion density, and $K_\text{cx}$ is the rate of charge-exchange as a function of $(x,v)$ and $\frak p$. Here, $s_\text{cx}$ is the distribution at which charge-exchange events occur, given by the solution of the linear system 
\be (\text{Id}-P_\sigma)s_\text{cx}=P_\sigma s_0.\ee

\noindent This latter approach is taken in \cite{umansky2025machine}. In general, (\ref{linsstar}) can be solved either by expanding the Neumann series directly, or by choosing a solver adapted to the structure of the matrix for the given parameters and geometry. For example, in systems where neutral mean-free-paths are short, $P_\sigma$ will generally be quite sparse, in which case sparse solver (e.g. Krylov) methods could reasonably outperform direct expansion of the Neumann series.

\subsubsection{Fixed-Point Iteration}
\label{fixedptit}
Because $\|P_\sigma\|<1$, Eq. (\ref{sstardef}) implies that $s_\star$ is the unique global fixed-point of the contraction mapping 
\be s \mapsto s_0 + P_\sigma s.\label{fixedpoint}\ee

\noindent Note that the first term is the {\it original} source $s_0$ from Eq. (\ref{Boltz}); in particular Eq. (\ref{fixedpoint}) is not linear. The fixed point iteration with initial guess of $s_0$ has iterates
$$s_M=s_0 + P_\sigma s_0 + P_\sigma^2 s_0 +\ldots + P_\sigma^M s_0=(\text{Id}+P_\sigma)^Ms_0,$$

\noindent which reproduces the Neumann series (\ref {sstardef}). More generally, however, iterating (\ref{fixedpoint}) from any initial guess will converge to $s_\star$, a property that is particularly useful for coupled simulations with dynamic plasma parameters (see Section \ref{expcoupling}). Although in our case Eq. (\ref{fixedpoint}) is affine, this method generalizes easily to non-linear cases where neutral-neutral interactions are accounted for.

We emphasize that, thus far, the linear algebra involves operators acting on functions on {\it phase space}. In particular, the resulting numerical implementations of these linear systems are of untenable size in practice, due to the ``curse of dimensionality''. The localization assumption (VLP) reduces these systems to manageable size, as we now describe.

\subsection{Constant Charge-Exchange Submanifolds}
\label{constantcxsub}

When the above scheme is employed using particular families of kernels indexed by charge-exchange, the resulting linear algebra systems collapse to ones whose size is determined by the {\it real space} dimension alone.

Let $J_\sigma$ and $K_\sigma$ be the families of kernels indexed by positive integers $\sigma$, defined as follows when acting on a particle density distribution $s$ : (i) $J_\sigma$ is the integral kernel for which $J_\sigma * s$ is the distribution of where particles are located after advancing the state by precisely $\sigma$ charge-exchange collisions\footnote{Since the source is assumed to be constant in time, the output of $J_\sigma*s$ updating the positions includes another copy of the original $s$ for the new particles, thus the identity factor in (\ref{sourcecorrection}).}, and (ii) $K_\sigma$ is the integral kernel for which $K_\sigma  * s$ is the distribution proportional to where particles have spent time when advancing the state by precisely $\sigma$ charge-exchange collisions. The expressions to calculate these quantities with estimators in a chosen discretization appear below in Eqns. (\ref{KsigmaT})--(\ref{jsigmaE}).

Suppose now that the localization assumption (VLP) holds, so that particles that charge-exchange at $x$ have outgoing velocity $v\sim \mathcal N_{\mathfrak p(x)}(v)$ chosen from a distribution determined by the local plasma parameters $\frak p(x)$. Thus, at the infinitessimal instant after charge-exchange, the velocity distribution is determined implicitly as a function of the spatial distribution. Convolution with $J_\sigma$ therefore has the form

\be  \label{compression}
J_\sigma(x,v,x',v';\mathfrak p) =\mathcal N_{\mathfrak p(x)}(v)\cdot j_{\sigma}(x,x';\frak p) 
\ee

\noindent provided the integer collision index $\sigma \neq 0$. Here, $j_\sigma(x,x';\frak p)$ is an integral kernel in only the {\it spatial} variables. More specifically, once $\sigma\geq 1$, the input distribution also has the form $s_1(x',v')=f(x')\mathcal N_{\frak p(x')}(v')$, and $J_\sigma$ acts by 

$$J_\sigma(x,v,x',v') * s_1=\mathcal N_{\frak p(x)}(v) \cdot j_\sigma(x,x') * \int_{\R^3} s_1(x',v')dv'$$

\noindent where the convolution $*$ on the right side only integrates over the $x'$ variables. 

Below, we will show that the particular family of kernels $J_\sigma$ and $K_\sigma$ advancing the particles by $\sigma$ resonant charge-exchange collisions (defined in Eqns. (\ref{KsigmaT})--(\ref{jsigmaE})) obey the generalized semi-group property (A2)--(C2). This allows the virtual source $s_\star$ to be calculated as $$s_\star(x,v)= N_{\frak p(x)}(v)\cdot \Big[\underbrace{j_\sigma * j_\sigma * \ldots *j_\sigma}_{M \text{ times}} * s_0\big ]$$

\noindent for sufficiently large $M$. Velocity information is needed only for the original distribution $s_0(x,v)$ and the final application of $K_\sigma$ in the setting that the equilibrium distribution of velocities is of interest. In particular, no intermediate velocity information is needed, and the linear algebra in Section \ref{multilevel} reduces to the dimension of the spatial indices only.

\medskip

\section{Collisional Propagators}
 
\label{MLMCSection}
The iterative methods to obtain the virtual source $s_\star$ in Eqs. (\ref{linsstar}) and (\ref{fixedpoint}) coincide when the former is solved by Neumann series, with the iterates being
\begin{eqnarray}
s_1&=& s_0 + P_\sigma s_0\nonumber\\
s_2&=& s_0 + P_\sigma s_1\nonumber \\
& \vdots &  \label{fixedpointit}\\ 
s_M&=& s_0 + P_\sigma s_{M-1}.\nonumber
\end{eqnarray}
In this section, we show these successive approximations fit into the standard framework of a multi-level Monte Carlo scheme. To begin, we first recall the language of Monte Carlo estimators.

\subsection{Discretization and Estimators}
\label{MCdescription}
Recall that a Monte Carlo method obtains an approximate solution to a discretization of Eq. (\ref{Boltz}) by generating sample particle flights or {\it trajectories}. In what follow, we use underlined variables e.g. $\underline n(z), \underline K(z,z')$ to denote the discretized approximations of the continuum objects in the previous subsection, though we conflate these for the spatial, velocity and plasma variables and continue to write $z=(x,v)$ and $\frak p$. Let $\mathcal P$ denote the high-dimensional\footnote{We assume there is a very large, user-input cap on the maximum number of collisions in possible trajectories, thus $\mathcal P$ is finite-dimensional.} ``pathspace'' $\mathcal P$ of possible trajectories. 
Sample trajectories $\gamma$ drawn from a probability distribution $d\Gamma_{\frak p,s_0}$ on $\mathcal P$. The subscript indicates that, while the space of trajectories is fixed, the probability distribution on it depends on the plasma parameter $\frak p$ and the starting distribution $s_0(x,v)$. An {\it estimator} is a mesh-valued random variable, $\mathcal T$ such that 
\bea
\underline n_\star(x,v)=\mathbb E(\mathcal T)=\int_{\mathcal P} \mathcal T(\gamma) \ d\Gamma_{p,s_0}(\gamma)\approx \frac{1}{N}\sum_{i=1}^N \mathcal T(\gamma_i)
\eea

\noindent where $D=O(h^{2d})$ 
is the number of volumes cells (for volume-cell diameter $h$ and spatial dimension $d$), and $N$ is the number of sample trajectories. For example, the track-length estimator $\mathcal T_{\text{TL}}$ integrates the distance the trajectories travels through each volume, whereas the collisional estimator $\mathcal T_{\text{coll}}(\gamma)$ is simply the basis-vector indicating the volume cell in $\R^D$ where $\gamma$ terminates. 

In this language, the (discretized) Green's function is 

$$\underline K(z,z';\frak p)=\int_{\mathcal P} \mathcal T(\gamma) \ d\Gamma_{\frak p,\delta(z')}(\gamma)\approx \frac{1}{N}\sum_{i=1}^N \mathcal T(\gamma_i)$$

\noindent i.e. the sum of the value of the estimator over $N$ sample trajectories with background plasma $\frak p$ and originating from a point source $\delta(z')$. The normal output of a Monte Carlo method is obtained by summing over trajectories with start points sampled from the source $s_0(x,v)$, corresponding to the discretized version of convolution Eq. (\ref{convolution}). 
\subsection{Defining Collisional Propagators}
\label{collisionalpropagators}
Each particle trajectory $\gamma$ undergoes a finite number of collisions $c(\gamma)$. For any $\sigma\in \N$, we denote the trajectory truncated at the $\sigma^{th}$ collision by $\pi_\sigma\gamma$. In this case, we define the following three quantities.  

\begin{eqnarray}
\underline K_\sigma(z,z';\frak p)&=&\int_{\mathcal P_{\sigma}} \mathcal T(\pi_\sigma \gamma)d\Gamma_{\frak p,\delta(z')}(\pi_\sigma\gamma) \nonumber \\ &=&\frac{1}{N}\sum_{i=1}^N \mathcal T(\pi_\sigma\gamma_i) \label{KsigmaT}\\
\underline P_\sigma(x,x';\frak p)&=&\int_{\mathcal P_{\sigma}} \mathcal E(\pi_{\sigma}\gamma)d\Gamma_{\frak p,\delta(z')}(\pi_\sigma\gamma)  \nonumber \\ &=&\frac{1}{N}\sum_{i=1}^N \mathcal E(\pi_\sigma\gamma_i)\label{ksigmaE} \\
\underline j_{M\sigma}(x,x';\frak p)&=& \sum_{n=0}^M \underline P_{n\sigma}(x,x';\mathfrak p),\label{jsigmaE}.
\end{eqnarray}
\noindent We also define $J_\sigma$ via Eq. ( \ref{compression}) using the above $j_\sigma$. In the above, $\mathcal T$ is a (phase-space valued) estimator of choice, and $\mathcal E$ is the (real-space valued) end-of-flight estimator, whose value is $1$ in the volume-cell where the $\sigma^{th}$ collision of $\gamma_i$ occurs and $0$ elsewhere.
Here, the dependence of the estimator on the mesh variable e.g. $\mathcal E(\pi_\sigma \gamma)=\mathcal E(\pi_\sigma \gamma)(x)$ is kept implicit in the notation. 
$d\Gamma_{\frak p,\delta(z')}$ is now, technically speaking, the pushforward probability measure on the space of truncated paths. Sample trajectories of length $\sigma$ are generated from this pushforward measure in the obvious way: by truncating samples from the full trajectory space $\mathcal P$. 

We now argue that with these definitions, the families of propagators $\underline K_\sigma, \underline J_\sigma$ obey the Properties (A2)--(C2) necessary for the multi-level iteration scheme of Section \ref{kernelssection} to apply. Establishing this shows that Eq. (\ref{linsstar}) or (\ref{fixedpoint}) correctly obtains the equilibrium distribution. Property (C2) is the definition of $J_\sigma$: the source of particles after $\sigma$ collisions, tracked for $\tau$ additional collisions gives precisely the tally after $\sigma+\tau$ collisions. Property (B2) is likewise straightforward: provided the ionization rate is non-zero, trajectories eventually terminate, and as $\sigma$ increases, the contribution to the approximate distribution $\underline n_\star(x,v)$ from trajectories longer than $\sigma$ approaches 0. In particular, for any finite sample size $N$, $\underline K_\sigma$ and $\underline K$ are eventually equal; this establishes (B2). For (A2), note first that $P_\sigma \circ P_\sigma=P_{\sigma+\tau}$ is immediate, simply because performing $\sigma+\tau$ collisions is the same regardless of whether intermediate information is tallied after the first $\sigma$. (A2) then follows from expanding (\ref{jsigmaE}) and using that $P^{k}_\sigma=P_{k\sigma}$.

\subsection{Multi-Level Monte Carlo}
\label{mlmc}

The iterative process taking $M\to \infty$ defines a multi-level Monte Carlo scheme \cite{giles2015multilevel}. For simplicity, we explain in the case that $\sigma=10$, and $\mathcal T_\text{coll}$ is a collisional estimator. The $N$ fixed sample trajectories with source $s_0$ can be subdivided into two groups: those that terminate before the $10^{th}$ collision, and those that do not. Proceeding in this fashion, the trajectories are divided into levels, with the $j^{th}$ level being the trajectories ionizing between the ${10j}^{th}$ and the $10(j + 1)^{th}$ collisions. Let the $j^{th}$ source $s_j$ be the distribution of phase-space positions of the particles at the $10j^{th}$ collision among those trajectories that have not terminated.

The levels then proceed as follows: the particles starting from $s_j$ are either ionized within $10$ collisions, in which case they contribute to $K_{10(j+1)}(z,z';\frak p) - K_{10j}(x,z';\frak p)$ OR they endure, in which case they contribute to $s_{j+1}$. By the same argument as for (B2), allowing the number of levels to tend to infinity eventually accounts for all trajectories. Thus the full source-correction operator as defined in Eq. (\ref{sourcecorrection}) is $$\underline P_{\sigma}(s)=\frac{1}{N}\sum_{\alpha=1}^N \mathcal E(\pi_{\sigma}\gamma_\alpha),$$
\noindent i.e. it is the end-of-flight position tracker for trajectories of length $\sigma$ with starting points drawn from $s$.

The successive approximation create a telescoping sum of corrections to define a multi-level scheme. Indeed, the argument above also implies that for any positive integer $\sigma$ collisionals,  

$$\underline K(z,z';p)=\underline K_{\sigma}(z,z';p) +  \sum_{j=2}^\infty (K_{\sigma j}(z,z';p)-K_{\sigma(j-1)}(z,z';p))$$
\noindent where 
$$(K_{nj}(z,z';p)-K_{n(j-1)}(z,z';p))\approx \frac{1}{N_j}\sum_{\alpha=1}^{N_j} \mathcal T_{coll}(\pi_{10}\gamma_\alpha)$$
\noindent and the latter sum is over $N_j$ sample trajectories of length $10$ with source $s_i$. Collapsing the telescoping sum, we see that $$\underline K(z,z';p)\approx \sum_{j=1}^\infty \frac{1}{N_j}\sum_{\alpha=1}^\infty \mathcal T_{\text{coll}}(\pi_{10}\gamma_\alpha),$$
or if all $N_j$ are equal then $\underline K(z,z';p)=\tfrac{1}{N}\sum_{\alpha=1}^N\mathcal T_{\text{coll}}(\pi_{10}\gamma_\alpha)$ where trajectories are sampled from $s_*=s_0+s_1+\ldots$. The same logic applies precisely with the track length estimator $\mathcal T_{TL}$ in place of the collisional estimator. In either case, this framework opens the possibility for the application of many techniques available for MLMC methods, such as judiciously choosing the number of samples $N_j$ at each level. 

\noindent

\bigskip

\section{Propagator-based MLMC} 
\label{PropagatorMLMC}

This section describes the numerical implementation of the MLMC scheme in Section \ref{mlmc}.

\subsection{Implementation}
It is straightforward to implement the propagator MLMC scheme from a pre-existing Monte Carlo code for kinetic neutral species. Two trackers are required: (i) an end-of-flight estimator $\mathcal E$, which histograms the locations of particles after a specified number $\sigma$ of collisions, and (ii) a standard estimator $\mathcal T$, usually a collisional or track-length estimator. Here, we employ a track-length estimator with attenuated absorption for $\mathcal T$ and henceforth restrict the discussion to this case.

Monte Carlo flights are run for particles \cite{heifetz1986neutral,wiesen2015new,DEGAS2manual} and terminated at collision $\sigma$; results tallied with the end-of-flight estimator $\mathcal E$ and the track-length estimator $\mathcal T_{\text{TL}}$ yield approximations for $\underline P_\sigma(\delta(z-z'))$ and $\underline K_\sigma(z,z';p)$ respectively as in Eqns. (\ref{KsigmaT})--(\ref{ksigmaE}). These propagators are combined as the columns of a matrix to form the full propagation operator $\underline P_\sigma$ and the full $\underline K_\sigma*$ matrix, from which it is straightforward to implement both the linear system (Eq. \ref{linsstar}) and fixed-point methods (Eq. \ref{fixedpointit}), and obtain the equilibrium via (\ref{finalapplication}). 

We note, before continuing, that reduced computation time is {\it not} among the advantages of propagator method, nor is this its goal. The statistical error of the output distribution is roughly proportional to the total number of collision events simulated, at least for a track-length estimator; thus although the particle trajectories for the propagator method are shorter, proportionately more must be run to obtain the same statistical error. The total computation time for the propagator method is therefore, in general, approximately equal to standard approaches.

\subsection{Sources of Error}
\label{errorsection}
There is an additional source of discretization error in propagator MLMC compared to standard (i.e. steady-state) MC. Indeed, $\mathcal E$ collects particle locations after testflights at level $j$ to form the source $s_{j}$; at the next stage, new particle flights are begun according to the distribution $s_{j}$ from the center of each volume cell.

This re-centering can create a bias in the distribution by introducing an artificial ``spreading'' effects, especially when the neutral distribution is steep and narrow in which case more particles tend to end their flights on the side of the volume cell towards the center of the distribution, and re-centering pushes them outward. This bias can become large if mean-free-paths are comparable to volume-cell widths (see also the Discussion \ref{refinements}). 

In addition to recentering the particles, each stage of the MLMC redistributes the particle weights so that all the weights in each cell are equal at the start of the next stage. Physically, this is a trivial operation: recall that in the attenuated absorption method each trajectories represents an ensemble of identical particles the number of which is given by the weight, thus reweighting is simply a relabeling of identical particles within a volume cell. From purely statistical arguments, however, it is far from obvious that the distribution of particle weights at the start of each new level does not affect the result. The numerical experiments in the next section confirm that this weight redistribution indeed does not disrupt finding the equilibrium .

\label{reproducingsection}

 \begin{figure*}[h!]
    \centering
    \includegraphics[width=\textwidth]{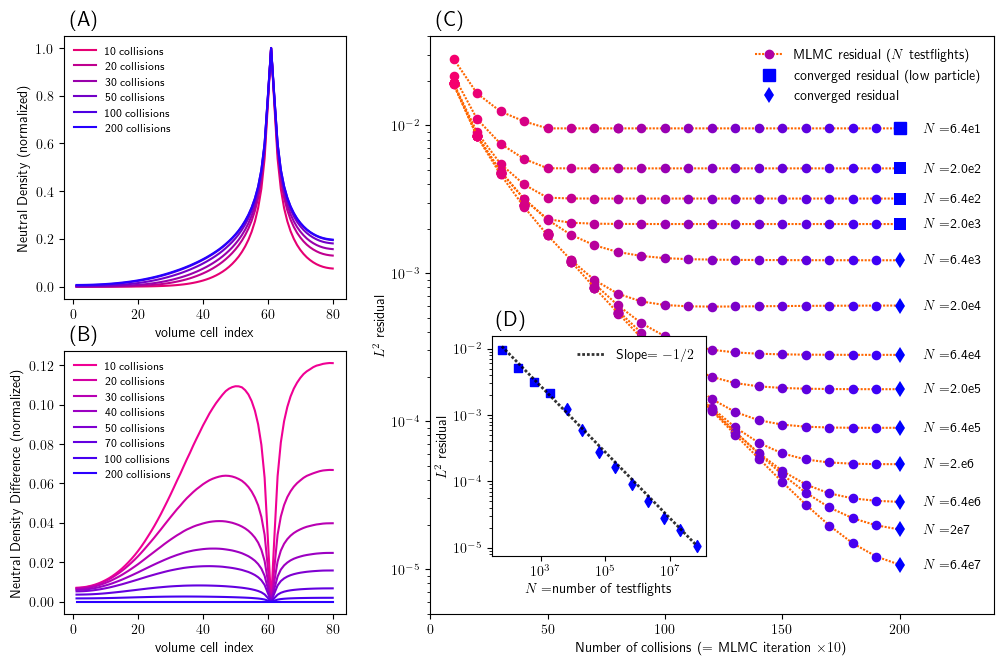}

    \caption{ (A) 1-dimensional slices of the 2-dimensional propagator profiles converging as collision number $\sigma\to \infty$ to the steady-state propagator $K_\infty$ (solid blue). (B) The difference of the same profiles from the limit $\underline K_\infty$. (C) The $L^2$-convergence of the propagator iteration as a function of number of collisions $\sigma$ for a variety of total particle trajectories $N$. Colors fading from magenta to dark blue mirror those in (A)-(B), and converged equilibria are indicated by squares or diamonds for each fixed $N$. (D) $L^2$ residual difference between the equilibrium obtained by propagator MLMC and the equilibrium obtained by standard MC, as a function of total flight number. Datapoints are the converged equilibria obtained in (C) indicated by the same markers. The expected convergence rate $O(N^{-1/2})$ is indicated by the dotted black line. }
   \label{Fig1}
\end{figure*}
\section{Propagator MLMC Reproduces Standard MC}
\label{reproducessection}
Sections \ref{kernelssection}--\ref{MLMCSection} showed by analytic and physical arguments that 
\begin{eqnarray} \xlim{M}{\infty}\|\underline K_{M\sigma}(x,x';\mathfrak p) - \underline K_{\infty}(x,x';\mathfrak p)\|_2&=&0   \label{B2proof}\\ \|\underline K_{MLMC}(x,x';\mathfrak p)-\underline K_{\rm{ref}}(x,x';\mathfrak p)\|_2&=&O(N^{-1/2}). \nonumber \\  \label{18proof}
\end{eqnarray}

\noindent where $\|\cdot \|_2$ is the $L^2$-norm on the spatial grid, and $\underline K(x,x')$ is obtained from $\underline K(z,z')$ by integrating over veloctiy variables. Eq. (\ref{B2proof}) is property (B2), and Eq. (\ref{18proof}) states that the discretized version of the steady-state Green's function obtained via the propagator MLMC method reproduces the true discretized Green's function (obtained by a chosen reference method) up to the statistical error, where $N$ is the number of trajectories sampled. Eq. (\ref{18proof}) implies the result holds for any source distribution, by linearity. 

In this section, we further evidence these claims by benchmarking a numerical implementation of the propagator MLMC scheme against a standard reference implementation of Monte Carlo. By ``standard'' we refer to a scheme in which single particle flights are run until ionization and statistics tallied with a track-length estimator.

\subsection{Numerical Setup} \label{setup} Numerical experiments were conducted in a reduced model with simplified geometry and atomic physics. The MLMC scheme is implemented numerically using an updated 2d-version of the code developed in \cite{parker2025coupling}. All simulations are run on rectilinear tokamak meshes; that is to say, the mesh includes branch cuts that make it into a tokamak topologically but it retains a straight (i.e. rectilinear) geometry. Only the spatial equilibrium distribution is tallied, so that the density in every mesh cell integrates over the velocity variables. These choices are simply for convenience of implementation and there should be no change to the methods described for complex geometries or resolving velocity profiles.

The outputs of Propagator MLMC were compared to those of standard MC for a variety of simulation and plasma parameters. Figure \ref{Fig1}(A--D) depicts data for the following plasma parameters: $\frak p$ is a steady-state plasma configurations for a simple non-linear diffusion model $ -\nabla (D(\frak p)\nabla \frak p)$ (with fixed constant temperature) with a source at the core-boundary and Dirichlet boundary conditions at the remaining boundaries. This background $\frak p$ yields a regime where neutrals undergo $\sim 100-200$ charge-exchanges before particle weights become negligible, and mean free paths are $O(1)-O(10)$ times the volume cell width, depending on the local plasma density. Simulations use the following numerical parameters: (i) a mesh of resolution $80 \times 80$, (ii) propagator iterations of $\sigma=10$ collisions, (iii) a point neutral source at a point $x'$ in the scrape-off-layer with grid coordinates $(i,j)=(60,40)$, (iv) the fixed point method Eq. (\ref{fixedpoint}) is used to solve Eq. (\ref{linsstar}), and (v) $\mathcal T$ is a track-length estimator with attenuated absorption.  

\subsection{Equilibrium Distribution}

Fig \ref{Fig1} (A) shows 1-dimensional slices of the family of 2-dimensional propagators $\underline K_\sigma(x,x';p)$ for various values of the collision numbers $\sigma$. As particle number increases, the profiles of converge and stabilizer on the limiting propagator $\underline K_\infty(x,x';\frak p)$. Fig \ref{Fig1} (B) shows the difference quantity (\ref{B2proof}) for the same family of propagators. As the level in the MLMC scheme increases, the result of the propagator iteration better approximates the tails of the distribution. 
In Fig \ref{Fig1}(A,B), the propagator profiles are normalized so that the maximum at the source point has unit value; in reality, since the sequence of sources $s_j\to s_\star$ is a cellwise increasing convergent sequence, the peak also grows in the unnormalized output. Together, Fig. \ref{Fig1})(A,B) numerically reaffirm property (B2) from Section \ref{semigroupprop} and Eq. \ref{B2proof}.  

Fig. \ref{Fig1}(C,D) show the result of increasing the number of particle trajectories $N$. A reference propagator $\underline K_{\text{ref}}(x,x';\frak p)$ is generated using the standard MC method with the maximum number of trajectories $N=2 \cdot 10^8$. Approximate propagators from the MLMC scheme are compared to this reference by the $L^2$ residual Eq. (\ref{18proof}). Fig \ref{Fig1}(C) shows the convergence of the $L^2$ residuals as a function of MLMC iterations for various testflight numbers $N$. For a fixed particle number, the $L^2$ residuals are plotted by circles fading from magenta to blue to mirror the corresponding profiles in (A,B). For any fixed particle number $N$, the propagator profiles for $\sigma=10,20,30,\ldots, 200$ convergence log-linearly for a period and then stabilize. The final $L^2$ residuals of the converged iteration are indicated by blue squares/diamonds, and decrease to zero with increasing number of trajectories $N$. Fig. \ref{Fig1}(D) The converged MLMC residuals from (C) plotted as a function of number of trajectories on a log-log scale. The dotted line indicates the analytic value of the log-linear slope $m=-1/2$ with best-fit vertical offset corresponding to the expected $O(N^{-1/2})$ convergence with number of testflights. The best-fit slope (not depicted) has a value of $m=-0.5076$. 

For $N<6400$ there are not enough testflights to distribute evenly across all volume cells. In this case, which cells particles start flights from is a user-input choice, and the result is highly dependent on this choice. For these low particle numbers, the residuals are demarked by a blue square rather than a blue diamond to distinguish such a choice has been made. 

\subsection{Rare-event Sampling}
A standard use-case of multi-level Monte Carlo methods is for rare event sampling. Here, as a second confirmationt that the propagator method is functioning as expected, we compare the ability of the propagator MLMC scheme to sample the rare events at the tails of distribution. 

\begin{figure}[h!]
    \centering
    \includegraphics[width=\columnwidth]{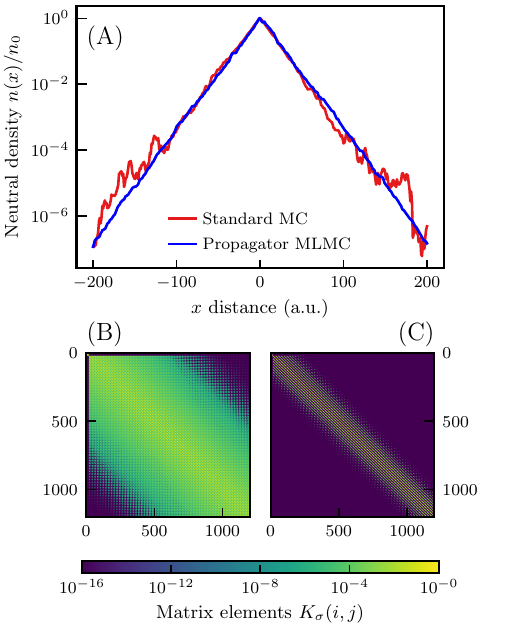}
    \caption{(A) Comparison of propagator-based MLMC and standard MC for sampling of tails in the fluid limit in one spatial dimension, where the analytic solution is given by Eq. (\ref{fluidlimit}).  (Bottom) Sparsity patterns of the matrices (B) $\underline K_\infty(x,x')$ and (C) $\underline K_{10}(x,x')$ for the numerical parameters used in Figure \ref{Fig1}. }
    \label{Fig3}
\end{figure}

Here, we take a simplified case in one spatial dimension setting where an analytic solution is known. The plasma background $\frak p$ is constant with temperature $T_0$ and ion density $u_0$, and we work in the fluid limit, i.e. the ratio $\sigma_{\text{cx}}/\sigma_{\text{iz}}$ of the cross sections is taken large ($\sim 100$), and values are scaled so that the mean free path is small compared to the volume cell diameter. In this setting, the spatial equilibrium distribution with source $s_0(x,v)=\delta(x-x_0)\mathcal N_{T_0}(v)$ is \be n_\star(x)=n_0 \cdot \text{exp}\left(\frac{|x-x_0|}{\ell}\right) , \label{fluidlimit}\ee
\noindent where $\mathcal N_{T_0}(v)$ is a Maxwellian at temperature $T_0$ and the exponent $\ell$ is determined by the plasma parameters and cross-sections. Fig. \ref{Fig3}(A) displays the comparison of the equilibrium obtained via the two methods on a high-resolution grid on a log-scale. The propagator method obtains log-linear decay from the source as predicted by Eq. (\ref{fluidlimit}) at far-removed spatial locations, whereas standard MC demonstrates large visible noise for moderate spatial separation from the source. 
 
There are several well-established techniques for similar rare-event sampling (e.g. modeling detector response) in kinetic neutral models, including russian-roulette sampling \cite{wiesen2015new} and adjoint Monte Carlo methods \cite{dekeyser2018divertor,baelmans2014efficient}. Although the impetus for introducing the propagator method is plasma-neutral coupling in the upcoming Section \ref{correlatedcoupled}, it may also provide a useful additional option for rare-event sampling. 

\subsection{Sparsity of the Propagator Matrix}
Provided that one is only concerned about the spatial density distribution at equilibrium, the steady-state propagator $\underline K_\infty$ can also be written as a matrix of the same dimensions as $\underline K_\sigma$. Because particle trajectories of lower collision numbers are far more localized in space than the full trajectories, the matrix representation of $\underline K_\sigma$ is generally significantly sparser than that of $\underline K_\infty$. Fig \ref{Fig3}(B,C) show schematics of the sparsity patterns of $\underline K_{10}$ for $\underline K_\infty$ for the same parameters as used for Fig. \ref{Fig1}. THese sparsity patterns depict the upper left corner of the $6400\times 6400$ matrices for clarity. In addition to its sparsity pattern, nearby columns of $\underline K_{\sigma}$ (or $\underline P_\sigma$) are in general almost translates of each other in slowly-varying plasma backgrounds, giving these matrices an ``Topelitz-like'' structure.  

This confirms that the propagator method indeed obtains the equilibrium distribution by repeated convolution of relatively localized updates. The Discussion (Section \ref{discussion}) analyzes the implications of this sparseness for applications of the propagator method in machine-learning based approximations.

\section{Coupled Plasma-Neutral Simulations using Propagator MLMC}
\label{correlatedcoupled}
\label{expcoupling}
As explained in the introduction, the primary motivation for the development of the propagator MLMC method is to improve correlation methods in coupled plasma-neutral models, with the ultimate goal being use in fully-implicit coupling by JFNK methods.

We consider a class of time-dependent coupled plasma- \ neutral models with state $(\frak p(t,x),n(t,x))$, where the two species interact through ionization, recombination, and additional processes such as wall recycling. A common simplification in many parameter regimes is the {\it quasi-steady-state assumption}, namely that the neutral population instantaneously equilibrates with respect to the current plasma state 
$\frak p$. We assume throughout that our system lies in such a regime. Under this assumption, the coupled system takes the schematic form
\begin{eqnarray}
\partial_t \mathfrak p + P(\mathfrak p) &=& s_1(n,\mathfrak p), \label{plasmaeq} \\
\mathcal B(n) &=& s_0(n,\mathfrak p), \label{neutraleq}
\end{eqnarray}
\noindent where $P(\frak p)$ is a continuum plasma mode, $\mathcal B(n)=v\cdot \nabla n + C_\frak p(n)$ is the time-independent Boltzmann equation for the neutral distribution $n$, and $s_0, s_1$ are now coupled source terms coming the interactions. The coupled system can be solved with two classes of methods, assuming a given solving method for each individual equation (\ref{plasmaeq}) and (\ref{neutraleq}). (i) Explicit coupling, e.g. the Lie-Trotter splitting method that alternatives updating $\frak p$ and $n$, substituting the previous update into the respective equations each iteration, or (ii) implicit coupling, in which the full system is treated as a single equation of the combined state $(\frak p,n)$ for which an implicit method is used.  When the explicit coupling scheme (i) is applied for the propagator method, solving Eq. (\ref{neutraleq}) to steady-state is replaced by a single application of the propagator in each iteration. As explained in the introduction, explicit methods suffer from slow convergence, whereas the instability of Jacobian-vector products prevents meaningful realization of the advantages of implicit coupling when (\ref{neutraleq}) is solved via Monte Carlo.

The real culprit behind slow convergence of explicit methods is the stiffness ratio. Multiscale physics and strong non-linearity force the stability threshold of the coupled system far below physical timescales.
Indeed, stability thresholds below $10^{-7}-10^{-6}$s for coupled simulations have observed in various settings \cite{boeyaert2024numerical}, whereas the macroscopic physical effects being modeled proceed orders of magnitude slower. This extreme stiffness indicates the potential for JFNK-based implicit methods to provide significant speed-ups, though these have only been successfully demonstrated in one spatial dimension, where the decorrelation problem seems to not appear \cite{parker2025coupling}. We refer the reader to \cite{parker2025coupling} for more detailed exposition of explicit vs. implicit coupling.

The most promising approach to obtaining stable Jacobian- \ vector products when using Monte Carlo is to use correlated Monte Carlo (CMC) methods\footnote{Often called ``colored noise'' in computational fluid dynamics.} For a well-chosen method of correlation, the MC output will be perfectly reproducible for fixed parameters, and vary in a smooth way with respect to varying parameters, allowing Jacobian-vector products to be calculated by finite-differences or auto-differentiation methods \cite{horsten2024AD,carli2023algorithmic}. Current correlation methods \cite{DEGAS2manual, ghoos2016accuracy} for standard Monte Carlo fall far short of the smoothness needed for JFNK methods to converge \cite{joseph2017coupling}. The failure of the correlation to allow stable Jacobian-vector products can be detected via the relative successive $L^2$ residual \be \text{res}_{L^2}(j):=\frac{\|\frak p((j+1)\Delta t,x) - \frak p(j\Delta t,x)\|_2}{\|\frak p(j\Delta t,x)\|_2}\label{successiveres}\ee between successive time-steps of the coupled system. The same sensitivity to changes in plasma background that prevents stable Jacobian-vector products also manifests as failure of the successive residual to converge to high precision. For a correlation method robust enough to achieve stable Jacobian vector products, this successive residual should drop to near machine precision. We therefore use (\ref{successiveres}) as a proxy measurement for the smoothness and robustness of correlation for the coupled system. 

We emphasize two points before displaying data: (i) the results displayed are for the convergence of {\it explicit} coupling methods; while the goal is to achieve implicit coupling, the purpose of the current study is to introduce the propagator method and demonstrate its validity and potential. Evaluation of the method for implicit coupling will be the subject of follow-up work. (ii) The successive $L^2$ residual (\ref{successiveres}) is a measure of the {\it precision} of the method, not of the accuracy. Even if the residuals of a correlated method converge to machine precision, numerical and statistical errors many orders of magnitude larger will, of course, still be present. Indeed correlation alone was shown to bestow no advantages for the purposes of accuracy \cite{ghoos2016accuracy}; our interest is simply that the high-degree of smoothness that is is a pre-requisite for the JFNK-based implicit methods also result in a high-degree of precision in the convergence.

\subsection{Propagator MLMC Demonstrates Robust Convergence}
\label{6.1}
\begin{figure}[h!]
    \centering
    \includegraphics[width=\columnwidth]{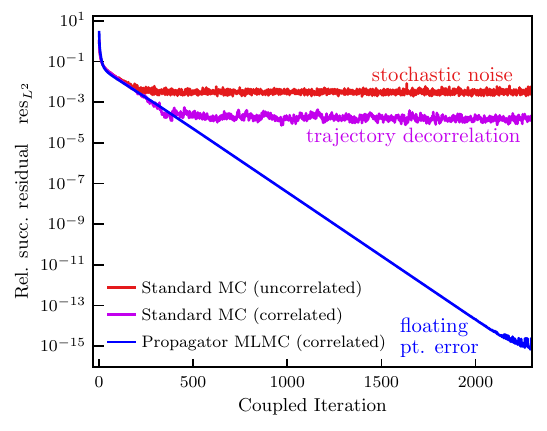}
    \caption{Relative successive $L^2$ residuals (Eq. \ref{successiveres}) for plasma-neutral simulation using explicit coupling. (Red) Standard MC residuals, without correlation. (Magenta) Standard MC residuals with current correlation methods. (Blue) Propagator MLMC residuals with correlation. }
    \label{Fig4}
\end{figure}

We find that our propagator MLMC method approaches an equilibrium of the coupled system with high precision: the relative successive $L^2$ residual defined by Eq. (\ref{successiveres}) drops to the machine precision level of about $10^{-15}$, whereas previous state-of-the-art methods only reach $10^{-3}-10^{-4}$ levels. To demonstrate this, we performed numerical experiments with coupled simulations solving Eqns. (\ref{plasmaeq}--\ref{neutraleq}) using three different Monte Carlo methods with the same simulation parameters, whose relative successive residuals are displayed in Fig. \ref{Fig4}: (i) standard MC (red), (ii) standard MC with current correlation methods (magenta), (iii) correlated propagator MLMC (blue).

For Fig. \ref{Fig4}, the choice of parameters was as follows. The mesh was a rectilinear tokamak geometry as in Section \ref{setup}, of resolution $40\times 40$. Coupling parameters were chosen to simulate perfect recycling coefficient at the device wall, and a simplified non-linear diffusion equation with $P(\frak p)$ in Eq. (\ref{plasmaeq}) given by $P(\frak p)=-\nabla(D(\frak p) \nabla \frak p)$ is used for the plasma model. Cross-section data was chosen to select a regime in which neutrals have high collisionality, and long mean-free-paths compared to device length-scales, this being the regime where current correlation methods were found to struggle the most. Additionally, the plasma flux to the device wall is recycling by a ``Maxwell's demon'' which deposits the scattered neutral particles as a point source directly in the scape-off-layer, to avoid confounding correlation with wall-screen effects.

The failure of the established correlation methods to produce stable Jacobian-vector products and smooth convergence of successive residuals (as in Fig. \ref{Fig4}, magenta curve) has been attributed to various properties of the dynamics of neutral particles in more than two dimensions.  Two speculative mechanisms for this are as follows: (i) The trajectories of two particles in infinitessimally different plasma backgrounds will diverge increasingly far over many collisions in 2-dimensions, whereas in a single dimension these differences tend to cancel out. (ii) Any opportunity for discrete transitions can cause particle histories to diverge if the parameters cause the same random number to fall on the other side of a cut-off value for triggering a certain event. To the author's knowledge, it has not been definitively established that these mechanisms are indeed the cause of the decorrelation for state-of-the-art methods; we also find it difficult to attribute the decorrelation of these methods to a single isolated cause in any particular simulation. Regardless of the mechanism causing the decorrelation, the propagator method (which was designed with these heuristics in mind) indeed prevents the same decorrelation.

\subsection{Stability with parameters}

  \begin{figure*}[h!]
    \centering
\begin{tikzpicture}
\begin{scope}[shift={(-0.0cm,0cm)}]

\node[inner sep=0] (right) at (1.5,-.8)
  {\includegraphics[width=0.36\linewidth]{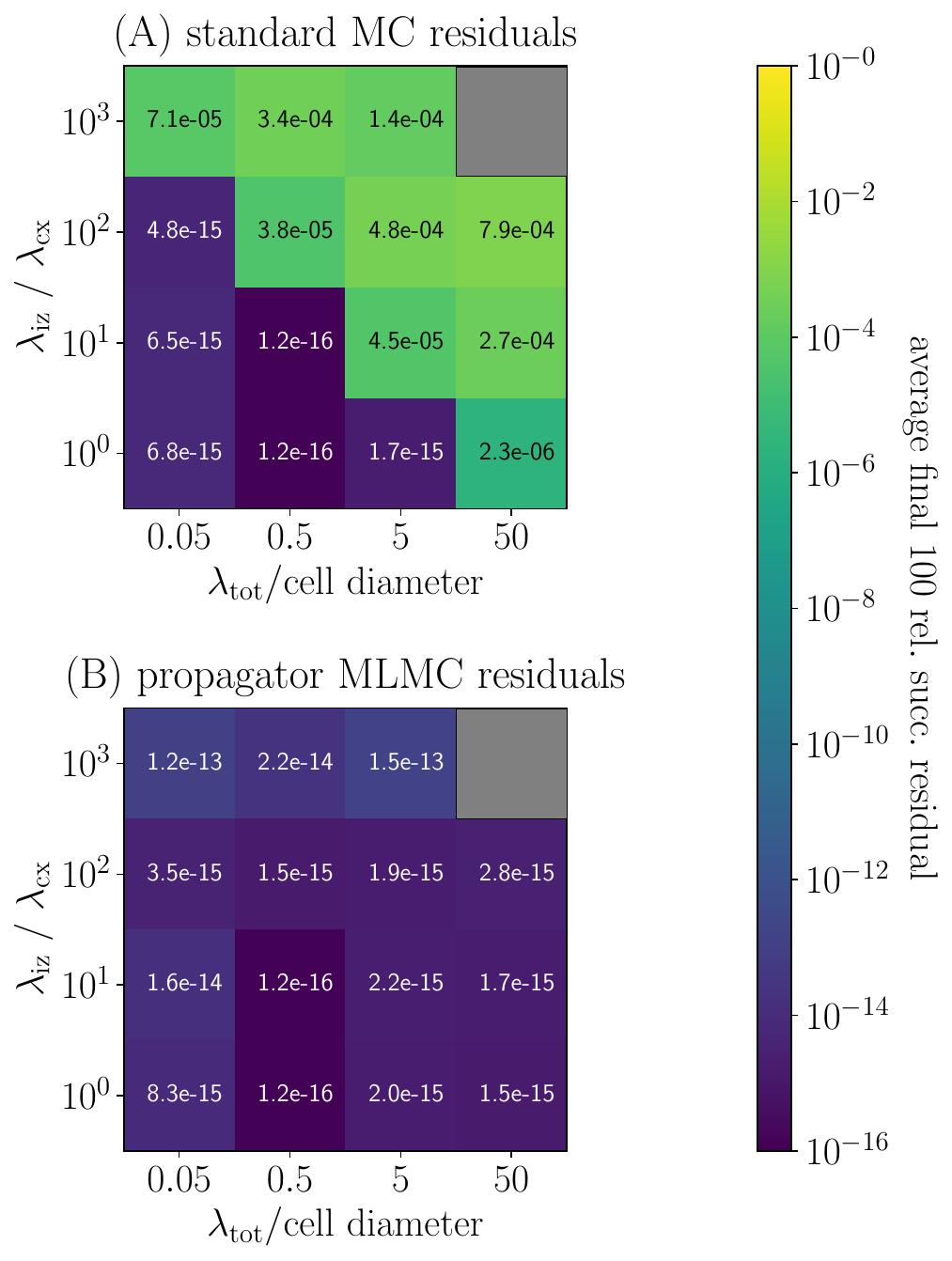}};


    \node[inner sep=4] (leftfig) at (-4.4,-.8)
  {\includegraphics[width=0.22\linewidth]{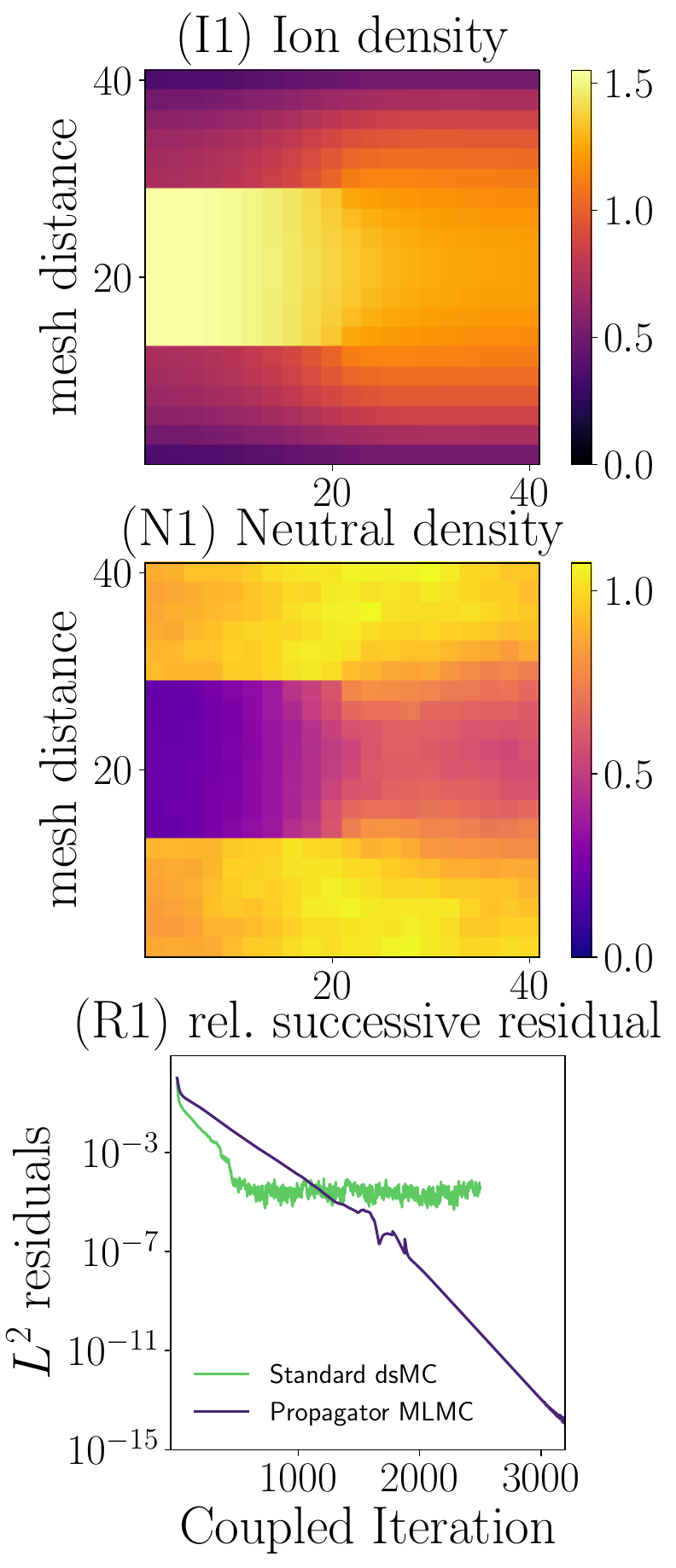}};

\draw[orange, very thick, dashed]
  (leftfig.south west) rectangle (leftfig.north east);

\draw[->, thick, orange] (leftfig) -- (0.3,-2.4);

    \node[inner sep=4] (leftfig2) at (7.4,-.8)
  {\includegraphics[width=0.22\linewidth]{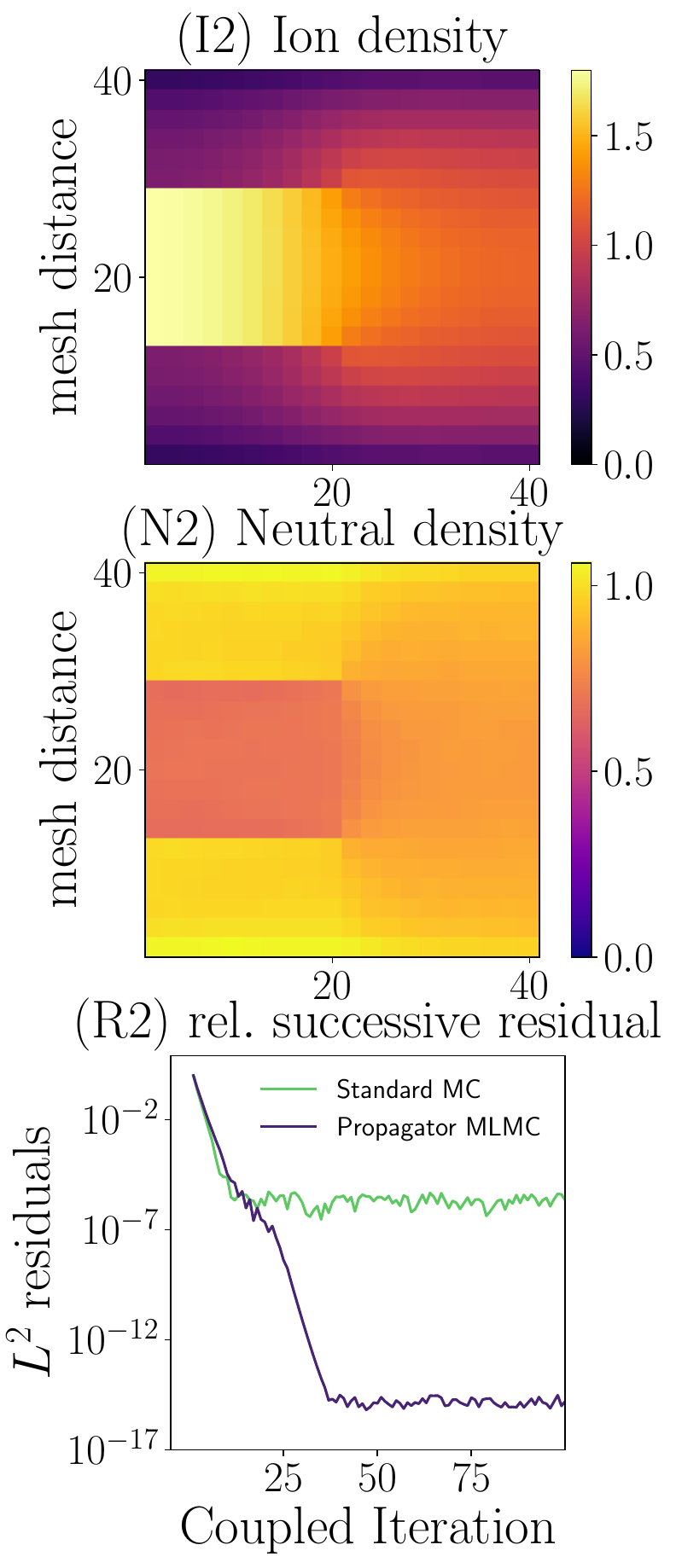}};
  \draw[orange, very thick, dashed]
  (leftfig2.south west) rectangle (leftfig2.north east);
  
\draw[->, thick, orange] (4.85+.5, -5.15) -- (1.9,-4.15);
    \end{scope}
\end{tikzpicture} 
    \caption{ (A) The successive residual for coupled simulations using standard MC and standard correlation methods as parameters vary. The average of the final 100 $L^2$-convergence residuals is displayed on a log scale heatmap. (B) The successive residual for coupled simulations using the Propagator MLMC method. (Insets, left and right) Coupled simulation plasma (I1, I2) and neutral (N1, N2) equilibrium state for two particular parameter pairs indicated by orange arrows, obtained by the propagator MLMC method (B). Equilibria for the two methods (A) and (B) have good visual agreement. (R1, R2) The relative successive residuals as a function of iteration for these two parameter pairs; colors match the final values in (A) and (B).}
   \label{parameterscan}
\end{figure*}
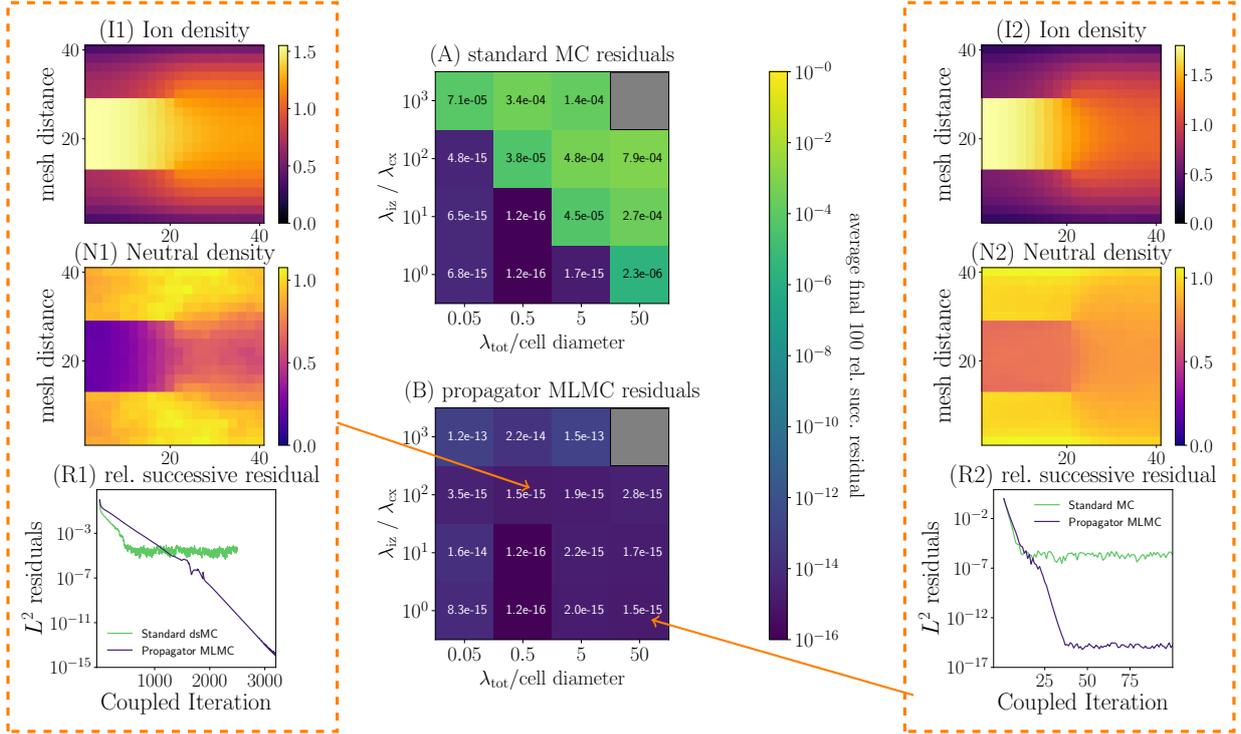
The robust convergence of the relative successive $L^2$-residual for the propagator MLMC scheme persists across all physical regimes tested. Figure \ref{parameterscan} below displays the final relative successive residuals (Eq. \ref{successiveres}) for the propagator MLMC scheme and standard correlated MC as physical parameters vary.

The two main physical parameters affecting the dynamics of neutral particles in this simplified model are (i) the ratio $\lambda_{\text{iz}}/\lambda_{\text{cx}}$ of the ionization mean-free-path to the charge-exchange mean free path, (ii) the ratio $\lambda_{\text{tot}}/L$ of the total mean free path to the cell diameter $L$ (thus length-scale of the device, since cell diameter is constant). These parameters are varied by directly adjusting the cross-sections and the mesh diameter respectively. Simulations were run for four different values of each parameter, giving a $4\times 4$ grid scanning the effective physical parameter space. For each pair of physical parameters, simulations were run to convergence for both standard MC with state-of-the-art correlation techniques, and propagator MLMC. These simulations were taken on reduced $20\times 20$ rectilinear meshes, with the same perfect wall-recycling and simplified non-linear diffusion plasma as in Figure \ref{Fig4}. Computation time is normalized by fixing the total number of collisions simulated per time-step across the two methods. Here, the plasma flux to the device walls were recycled as neutrals with source uniformly distributed along the walls.

Figure \ref{parameterscan} displays the final values of the relative successive residual (Eq. \ref{successiveres}) on a logarithmic scale for both methods for each pair of physical parameters. The values depicted are obtained by averaging the final $100$ values after the residual graph becomes flat.  Current correlation methods for standard MC demonstrate robust correlation for low-collision regimes (bottom region) where mean-free-paths are short compared to volume cell diameter (left region). For highly collision regimes with both short (top left, i.e. the near-fluid regime) and long mean free path (top right), the standard correlated MC residuals converge only a few orders of magnitude. This corroborates the heuristic that longer mean-free-paths and higher collisionality challenge standard correlation methods. In contrast, the propagator method is insensitive to varying physical parameters and achieves near machine-precision in all regimes, provided the time-step and value of $\sigma$ are chosen appropriately (values of $\sigma$ range from $1$ to $10$). At times, the propagator method relative successive residual is caught in a periodic pattern at values around $10^{-7}$, which we attribute to numerical resonance. The latter is a known side-effect of explicit coupling \cite{schlick1998nonlinear,ropp2005stability}, and it is unclear if this effect is meaningful for the purpose of Jacobian vector products. The top right parameter pair in the $4\times 4$ grid is omitted due to extreme computational costs: particles here undergo tens of thousands of collisions before ionization, with mean free path many times the domain diameter and each flight is costly to simulate; moreover, the dynamics force extremely short time-steps for explicit methods in this regime.

\section{Discussion}
\label{discussion}
This article introduced and verified a new multi-level Monte Carlo scheme for kinetic neutral modeling in edge plasma. The multi-level scheme, dubbed propagator MLMC, is based on iterative applications of event-based propagation operators. The method is applicable in the presence of common velocity-localizing collisions, assumption (VLP), which are resonant charge-exchanges in the case of neutrals in edge plasmas. In this situation, the evolution data of the system can be compressed into a matrix of the real space dimension, and the equilibrium obtained via
linear algebra on this matrix. 

The numerical experiments in Sections \ref{reproducessection}--\ref{correlatedcoupled} indicate promising preliminary results for applications of the propagator method in coupled plasma-neutral simulations. Below, we discuss the implications of the propagator method for future work in neutral modeling and related problems.

\subsection{Refinements Propagator MLMC}
\label{refinements}
The veresion of Propagator MLMC introduced above is effectively a first-order scheme in both real-space and velocity. The method could be extended to implement higher-order schemes in both, which would improve the method's accuracy and domain of applicability. 

\begin{enumerate}
\item[(i)] {\it (Higher-order in space)} At each iteration, the particles are recollected and flights for the next iteration restarted from the center of each cell. This can lead to a linear discretization error as explained in Section \ref{errorsection}. By expanding the neutral state-vector to also include $N$ higher-order moments of the distribution of collision locations within each cell after each iteration, this error could likely be reduced to $N^{th}$ order in the cell diameter. 

\item[(ii)] {\it (Higher-order in velocity)} The neutral state-vector could also be expanded to include $N$ higher moments of the distribution of incident {\it velocities} in each cell after each iteration. This could relax the assumption (VLP) to be that the outgoing velocity distribution is well-approximated to by a distribution determined by background field {\it and} the first $N$ moments of the incident distribution. This approach would also open the door to applications in regimes where (VLP) does not hold precisely, such as very cold detached divertor conditions.

\item[(iii)] {\it (Additional collision types)} The method was demonstrated for a simplified atomic physics model only account for charge-exchange and ionization effects. Extensions to include additional types of collisions and atomic physics interactions should be straightforward; for this we emphasize the propagator $K_\sigma$ advances by $\sigma$ resonant charge-exchanges, and multiple collisions of other types may occur in the intermediate flight with little change. The implications for decorrelation of including such additional interactions will be discussed in future work.  
\end{enumerate}

\subsection{Domains of applicability}
The distinguishing feature of neutral dynamics in edge plasmas is that Assumption (VLP) holds to a high degree of accuracy for resonant charge exchanges, and these are generally the dominant type of collision. To elaborate, below mid-level energies e.g. ($<\ \sim\!10$ eV), the angular cross-section is isotropic to high accuracy, before developing a forward-peak at the incoming velocity for extremely high energy neutrals. In contrast, the perfect “momentum swap” assumption (i.e. that the outgoing particle takes on the velocity of a thermal plasma ion) is valid for energies larger than about $\sim 0.01-0.1$ eV, below which it begin to develop anisotropy larger than $O(10\%)$ \cite[Sec. 2.11]{DEGAS2manual}. Many studies and code suites (e.g. EIRENE, DEGAS1) assume isotropic charge-exchange cross-section, though others (DEGAS2) implement table lookups to account for the deviation at very low eV. Thus, for neutrals in edge plasmas the assumption (VLP) would be challenged in cold, deeply-detatched divertor conditions. The higher-order methods proposed above could potentially be used to relax (VLP) and extend the method to this domain.

Versions of the kinetic Boltzmann equation where a version of (VLP) holds are ubiquitous in certain regimes of other physical domains besides neutral transport. In photon transport problems in edge plasmas, a version of (VLP) holds for photon scattering events in many situations \cite{rosato2016hybrid}, and likewise for regions of inertial confinement fusion plasmas in regions of high optical depth. The propagator method described here may also have use cases in these domains as well. 

There are, nevertheless, domains such as neutron and electron transport problems, with regimes in which obtaining the outgoing velocity distribution after collision events requires resolving the incident velocity distribution to high accuracy, not just a few moments. It appears unlikely to the authors that an extension of the propagator method to these situations is possible. Indeed, the validity of assumption (VLP) makes neutral particle transport in edge plasmas a simpler problem, by certain metrics, than its intellectual predecessor neutron transport. The propagator method provides one way  on this simplification.

\subsection{Fully Implicit JFNK-based methods}
As explained in the introduction, JFNK-based fully implicit methods have the potential to reduce the costs of coupled plasma-neutral simulations by orders of magnitude. Propagator MLMC offers a revised approach to coupled plasma-neutral simulations that addresses the key barrier: decorrelation from branching trajectories\footnote{Cf. the discussion in Section \ref{6.1}}. 

The results in Section \ref{correlatedcoupled} demonstrate that the propagator method effectively overcomes the decorrelation problem as it manifests in explicit coupling schemes. One heuristic explanation for this is as follow: the propagator method prevents diverging trajectories by mandating, by fiat, that all sampled trajectories considered are truncated a fixed small length $\sigma$, thus divergences simply do not have time to occur. The full multi-level scheme then samples the collection of full trajectories obtained by stitching these short trajectories end-to-end in all possible (exponentially many) combinations. Thus the correlation in the propagator method manifests, not as repeated sequences of random numbers, but as repeated motifs of short length $\sigma$ sub-trajectories that are shared across many full trajectories. 

Investigating the use of the propagator method for fully-implicit JFNK-based methods in more than one spatial dimension is, of course, the natural immediate follow-up to this work.

\subsection{Machine-learning based approximations}

Propagator MLMC may also offer avenues to accelerate machine-learning  methods for neutral particle dynamics. 
Machine-learning (ML) approaches based on various neural-network architectures have recently achieved success in approximating solutions of various PDEs and other problems in fusion \cite{jalalvand2025multimodal, degrave2022magnetic}. When exploring potential applications of these methods to kinetic neutral modeling, it is highly desirable to take advantage of any means of minimizing the computational load of training such neural network. 

The reduced sparsity pattern and Topelitz-like structure of the propagator matrix displayed in Figure \ref{Fig3} could be advantageous for developing neural-network approximations for kinetic neutral solvers. Indeed, reduced sparsity structure is generally considered an indicator that a neural-network approximation requires fewer parameter and thus a small training dataset. For networks trained on simulation data, the generation of large training datasets can constitute a significant portion of the cost. The authors believe that, because of the relatively large error display by many neural-network in related applications \cite{carey2025data,zhang2025calculation}, the most promising applications of such networks are in hybrid methods, using these networks in conjunction with the JFNK-based implicit methods above. Indeed, the low cost but suboptimal accuracy of neural networks in this domain make them ideal for use in warm-starts and preconditioning.  Neural-network approximations of the non-linear mapping $\frak p \mapsto \underline P_\sigma(x,x'; \frak p)$ taking the plasma parameter to the propagator matrix will be the subject of future work.

\section{Conclusions}

This article introduced a new multilevel Monte Carlo scheme for neutral particle transport that explicitly exploits a key structural property of neutral dynamics: the presence of frequent collisions with a velocity-localizing effect. The proposed multi-level scheme allows the equilibrium neutral distribution to be computed via linear algebra on sparse ``propagator'' matrices of the real-space grid dimension. It was verified that this method reproduces the results of standard Monte Carlo solvers in classical settings, but that it bestows several key advantages that make it a promising candidate for use in next-generation plasma-neutral coupling schemes. Namely, by enabling a new approach to trajectory correlation based on stitching together correlated shoft-trajectory motifs, the multi-level scheme overcomes the well-known limitation of rapid decorrelation caused by branching trajectories (see Section~\ref{6.1}).

This more powerful form of correlation makes the proposed method a natural candidate for use in fully implicit coupling methods for plasma-neutral systems based on Jacobian-free Newton-Krylov solvers -- a class of methods that has remained largely inaccessible due to decorrelation preventing stable Jacobian-vector products. Access to such solvers has the potential to dramatically accelerate coupled plasma-neutral edge simulations, while simultaneously mitigating known sources of error such as poor particle conservation. The sparsity structure of the propagator matrices also suggests a pathway to more accessible learned surrogate surrogate models for approximating neutral transport. In particular, the simpler structure of the propagator matrices may reduce the training load for neural networks designed to produce approximate neutral distributions; easier training of such networks could accelerate future investigations of their feasibility in neutral edge modeling.

\section{Acknowledgments}
G.P. is supported by an NSF Mathematical Sciences Postdoctoral Research Fellowship
(Award No. 2303102). Performed in part by LLNL under Contract DE-AC52-07NA27344. 
\bibliographystyle{amsplain}
{\small 
\bibliography{Bib_Statement2.bib}}
\end{document}